\documentstyle[prb,aps,epsf]{revtex}

\begin{document}

\draft
\title{Dynamical spin correlations in Heisenberg ladder under magnetic
field \\ and correlation functions in SO(5) ladder}
\author{Akira Furusaki$^*$ and Shou-Cheng Zhang}
\address{Department of Physics, Stanford University, Stanford,
California 94305-4060}
\date{July 29, 1998}
\maketitle
\begin{abstract}
The zero-temperature dynamical spin-spin correlation functions are
calculated for the spin-$\frac{1}{2}$ two-leg antiferromagnetic Heisenberg
ladder in a magnetic field above the lower critical field $H_{c1}$.
The dynamical structure factors are calculated which exhibit both
massless modes and massive excitations.
These modes appear in different sectors characterized by the parity in the
rung direction and by the momentum in the direction of the chains.
The structure factors have power-law singularities at the lower edges of
their support.
The results are also applicable to spin-1 Heisenberg chain.
The implications are briefly discussed for the various correlation
functions and the $\pi$-resonance in the SO(5) symmetric ladder model. 
\end{abstract}

\section{Introduction}

Systems locating in between one dimension and two dimensions, ladders,
have attracted great attention both theoretically and
experimentally.\cite{Dagotto} 
This line of research has been pursued both with the hope of finding some
clues to the understanding of the high-temperature superconductivity, and
because of the experimental discoveries of new ladder compounds,
such as a cuprate spin ladder\cite{Azuma} SrCu$_2$O$_3$ and a
superconducting ladder\cite{Uehara}
Sr$_x$Ca$_{14-x}$Cu$_{24}$O$_{41}$.
Another spin ladder system of current interest is the organic material
Cu$_2$(C$_5$H$_{12}$N$_2$)$_2$Cl$_4$ whose spin-gap behavior was
observed by the measurements of the spin susceptibility, the magnetization
curve, and the NMR relaxation rate.\cite{Chaboussant,Hammar} 
The nonzero magnetization is observed once the external magnetic field $H$
exceeds the lower critical field $H_{c1}$ which is equal to the spin gap
(in units where the product of the Bohr magneton and the $g$-factor is
set to be unity).
This is a quantum phase transition driven by the external magnetic
field from the gapped spin liquid to the gapless Tomonaga-Luttinger
liquid state. 
These experiments motivated further theoretical works on the various
properties of the gapless regime: the magnetization
process,\cite{Hayward,Cabra} the spin-spin correlation
functions,\cite{Chitra,Eder} and the spin-Peierls
instability.\cite{Nagaosa,Calemczuk}

In this paper we extend the theory of Shelton {\it et al.}\cite{Shelton}
to the two-leg Heisenberg ladder in the gapless regime ($H>H_{c1}$)
and calculate the dynamical spin-spin correlation functions and
structure factors in the ground state.\cite{Sachdev}
We obtain the dynamical structure factors containing massive spin
excitations as well as massless excitations.
The latter contribution is characteristic of the Tomonaga-Luttinger
liquid and is commonly found in the $S=\frac{1}{2}$ Heisenberg
chain.\cite{Muller,Talstra}
This component may be interpreted as coming from the bose-condensate of
$S^z=1$ magnons.\cite{Sachdev}
On the other hand, the massive excitations we are interested in have
$S^z=0$ magnons in their origin.
Below $H_{c1}$ the massive magnons give rise to $\delta$-function peaks in
the dynamical structure factor.
In the gapless regime ($H>H_{c1}$) the $\delta$-function peaks of the
massive excitations turn into power-law singularities, because the
massless excitations introduce algebraically decaying prefactors to
the exponentially decaying correlation functions.
An interesting feature of the antiferromagnetically coupled Heisenberg
ladder is that these massless and massive modes appear in different
sectors of the structure factors, characterized by the momentum $q_y$ in
the direction of the rungs ($q_y$ can take only 0 and $\pi$) and by the
momentum $q$ in the direction of the chains.

Our study of the massive excitations in the gapless regime is also
motivated to better understand the
$\pi$-resonance mode\cite{Zhang} in the SO(5) symmetric ladder model
introduced recently by Scalapino {\it et al}.\cite{Scalapino} 
In this model there is a quantum phase transition driven by the chemical
potential from the spin-gap Mott insulator to the $d$-wave-like
superconducting phase. 
It was shown in the strong-coupling limit that the spin-gap magnon
mode of the Mott insulator evolves continuously into the
$\pi$-resonance mode of the superconducting phase.\cite{Scalapino}
A complementary weak-coupling approach\cite{Senechal,Arrigoni,Lin} was
taken to study a general two-leg ladder model of
weakly interacting electrons.
In particular, Lin {\it et al.}\cite{Lin} showed that the model is
renormalized to a fixed point where a global SO(8) symmetry is
realized that contains the SO(5).
They also obtained the ground-state phase diagram and low-energy
excitation spectra at half filling which are qualitatively in
agreement with the strong-coupling picture of the SO(5) symmetric
ladder model. 
Yet the energy and the spectral weight of the $\pi$-resonance mode in
the dynamical spin structure factor are not completely understood even
for the ideally constructed SO(5) symmetric ladder.
An extensive numerical exact-diagonalization study is performed to
clarify this and related issues,\cite{Eder2} but clearly it is
desirable to develop an analytic theory to discuss the power-law
singularity which is expected to appear in the structure
factor.\cite{Scalapino} 
The field-driven quantum phase transition in the Heisenberg
ladder we study in this paper is analogous to the quantum phase
transition in the SO(5) symmetric ladder.
A $S^z=1$ boson in the Heisenberg ladder corresponds to a hole pair
in the latter model, and a $S^z=0$ boson to a magnon or the
$\pi$-resonance mode.
Therefore our Heisenberg spin ladder may be viewed as a toy model for
the SO(5) symmetric ladder.
Using the analogy, we can deduce the correlation functions for the
$d$-wave superconductivity, the charge density wave, and the spin
correlations in the SO(5) ladder from the spin-spin correlators of the
Heisenberg ladder.

This paper is organized as follows.
In Sec.\ II we introduce the model and analyze it in the limit of strong
interchain coupling, where it is easier to get intuitive pictures of the
physics. 
Detailed calculation of the dynamical spin-spin correlation functions and
the structure factors are presented in Sec.\ III.
Here we take the opposite limit of the weak interchain coupling, and
employ the bosonization method.
In Sec.\ IV we briefly discuss implications to the $\pi$-resonance in the
SO(5) symmetric ladder model and summarize the results.

\section{Strong-coupling approach}

The Hamiltonian for the two-leg Heisenberg ladder we study in this paper
is 
\begin{equation}
{\cal H}=
J\sum_{\mu=1,2}\sum^\infty_{i=-\infty}
\mbox{\boldmath$S$}_{\mu,i}\cdot\mbox{\boldmath$S$}_{\mu,i+1}
+J_\perp\sum^\infty_{i=-\infty}
\mbox{\boldmath$S$}_{1,i}\cdot\mbox{\boldmath$S$}_{2,i}
-H\sum_{\mu=1,2}\sum^\infty_{i=-\infty}S^z_{\mu,i},
\label{Hamiltonian}
\end{equation}
where $\mbox{\boldmath$S$}_{\mu,i}$ is spin-$\frac{1}{2}$ operator, and
the intrachain coupling $J$ is positive (antiferromagnetic).
The interchain coupling $J_\perp$ is also
assumed to be positive, unless otherwise noted.

In this section we briefly discuss static correlations above the lower
critical field $H_{c1}$ in the strong-coupling limit, $J_\perp\gg J$.
When $J/J_\perp=0$, the ladder is decomposed into independent rungs,
each rung consisting of two spins $\mbox{\boldmath$S$}_{1,i}$ and
$\mbox{\boldmath$S$}_{2,i}$. 
There are four eigenstates in each rung, the singlet
$|S_i\rangle=(|\!\!\uparrow\downarrow\rangle
             -|\!\!\downarrow\uparrow\rangle)/\sqrt{2}$,
and the triplet states
$|T_{i,+}\rangle=|\!\uparrow\uparrow\rangle$,
$|T_{i,0}\rangle=(|\!\uparrow\downarrow\rangle
                  +|\!\downarrow\uparrow\rangle)/\sqrt{2}$,
and $|T_{i-}\rangle=|\!\downarrow\downarrow\rangle$.
Their energies are $E(S_i)=-3J_\perp/4$, $E(T_{i,+})=J_\perp/4-H$,
$E(T_{i,0})=J_\perp/4$, and $E(T_{i,-})=J_\perp/4+H$, respectively.
It is then natural to introduce boson operators
$s_i^\dagger$, $t_{i,+}^\dagger$, $t_{i,0}^\dagger$, and
$t_{i,-}^\dagger$, which create the singlet state and the triplet states
at the $i$th rung.\cite{SachdevBhatt,Gopalan,Eder}
They obey the constraint,
$s^\dagger_is_i+t^\dagger_{i,+}t_{i,+}+t^\dagger_{i,0}t_{i,0}
 +t^\dagger_{i,-}t_{i,-}=1$.
With these hard-core bosons we represent the spin operators 
\begin{mathletters}
\begin{eqnarray}
S^z_{0,i}&\equiv&S^z_{1,i}+S^z_{2,i}
=t^\dagger_{i,+}t_{i,+}-t^\dagger_{i,-}t_{i,-},
\label{S^z_0,i}\\
S^z_{\pi,i}&\equiv&S^z_{1,i}-S^z_{2,i}
=t^\dagger_{i,0}s_i+s^\dagger_it_{i,0},\\
S^+_{0,i}&\equiv&S^+_{1,i}+S^+_{2,i}
=\sqrt{2}\left(t^\dagger_{i,+}t_{i,0}+t^\dagger_{i,0}t_{i,-}\right),
\label{S^+_0,i}\\
S^+_{\pi,i}&\equiv&S^+_{1,i}-S^+_{2,i}
=\sqrt{2}\left(-t^\dagger_{i,+}s_i+s^\dagger_it_{i,-}\right),\\
S^-_{0,i}&\equiv&S^-_{1,i}+S^-_{2,i}
=\sqrt{2}\left(t^\dagger_{i,-}t_{i,0}+t^\dagger_{i,0}t_{i,+}\right),
\label{S^-_0,i}\\
S^-_{\pi,i}&\equiv&S^-_{1,i}-S^-_{2,i}
=\sqrt{2}\left(t^\dagger_{i,-}s_i-s^\dagger_it_{i,+}\right).
\label{S^-_pi,i}
\end{eqnarray}
\end{mathletters}
The suffixes $0$ and $\pi$ refer to the momentum $q_y$.
Note that $S^z_{0,i}$ and $S^\pm_{\pi,i}$ do not contain the
$t_{i,0}$ boson.

In the presence of a high magnetic field $H\gtrsim J_\perp$, we may ignore
the $t_{i,0}$ and $t_{i,-}$ bosons as they have higher energies than the
$t_{i,+}$ boson. 
In lowest order in $J$ the effective Hamiltonian thus becomes
\begin{equation}
{\cal H}_{\rm eff}=
\frac{J}{2}\sum_i\left(
t^\dagger_{i,+}s_is^\dagger_{i+1}t_{i+1,+}
+t^\dagger_{i+1,+}s_{i+1}s^\dagger_it_{i,+}
+t^\dagger_{i,+}t_{i,+}t^\dagger_{i+1,+}t_{i+1,+}
\right)
+(J_\perp-H)\sum_it^\dagger_{i,+}t_{i,+},
\label{H_eff}
\end{equation}
with the simplified constraint $s^\dagger_is_i+t^\dagger_{i,+}t_{i,+}=1$.
As pointed out in Ref.\ \onlinecite{Mila}, this effective Hamiltonian may
be written as
\begin{equation}
{\cal H}_{\rm eff}=
J\sum_i\left(
\widetilde{S}^x_i\widetilde{S}^x_{i+1}
+\widetilde{S}^y_i\widetilde{S}^y_{i+1}
+\frac{1}{2}\widetilde{S}^z_i\widetilde{S}^z_{i+1}\right)
+\left(J_\perp+\frac{J}{2}-H\right)\sum_i\widetilde{S}^z_i+{\rm const},
\label{H_XXZ}
\end{equation}
where $\widetilde{S}_i$ is a spin-$\frac{1}{2}$ operator
\begin{mathletters}
\begin{eqnarray}
\widetilde{S}^z_i&=&t^\dagger_{i,+}t_{i,+}-\frac{1}{2},\\
\widetilde{S}^+_i&=&t^\dagger_{i,+}s_i,\\
\widetilde{S}^-_i&=&s^\dagger_it_{i,+}.
\end{eqnarray}
\end{mathletters}\noindent
We notice that Eq.\ (\ref{H_XXZ}) is just the Hamiltonian of the
spin-$\frac{1}{2}$ $XXZ$ chain with the external magnetic field
$\widetilde{H}=J_\perp+\frac{J}{2}-H$, whose properties are well
understood. 
When $|\widetilde{H}|<\frac{3}{2}J$, the $XXZ$ chain is not fully
polarized.\cite{Yangs} 
This means that the ladder has unsaturated magnetization
$\langle S^z_{0,i}\rangle=\langle t^\dagger_{i,+}t_{i,+}\rangle=M$
($0<M<1$) when $J_\perp-J<H<J_\perp+2J$.
We have thus obtained the lower and upper critical fields
$H_{c1}=J_\perp-J$ and $H_{c2}=J_\perp+2J$ in lowest order in
$J/J_\perp$.\cite{higher} 
The magnetization curve for $M\ll1$ can also be obtained from the
known result for the $XXZ$ chain:\cite{Yangs}
\begin{equation}
M=\frac{1}{2}+\langle\widetilde{S}^z_i\rangle=
\sqrt{\frac{2(H-H_{c1})}{\pi^2J}}.
\label{M}
\end{equation}
A similar result holds for $1-M\ll1$.
The square-root behavior is a well-known universal
behavior.\cite{Dzhaparidze,Tsvelik,AffleckBose,SakaiPRB,Chitra} 
When $0<M<1$ the ground state is a superfluid of the $t_{i,+}$
bosons.
Within our approximation the following equal-time spin correlation
functions are readily obtained from those
of the $XXZ$ chain:\cite{Bogoliubov}
\begin{mathletters}
\begin{eqnarray}
\langle S^z_{0,n+1}S^z_{0,1}\rangle&=&
\langle(\widetilde{S}^z_{n+1}+1/2)
(\widetilde{S}^z_1+1/2)\rangle=
M^2-\frac{c_1}{n^2}
+\frac{c_2}{n^\eta}\cos(2\pi Mn),
\label{static<S^z_0S^z_0>}\\
\langle S^x_{\pi,n+1}S^x_{\pi,1}\rangle&=&
2\langle\widetilde{S}^x_{n+1}\widetilde{S}^x_1\rangle=
c_3\frac{(-1)^n}{n^{1/\eta}}
-c_4\frac{(-1)^n}{n^{\eta+1/\eta}}\cos(2\pi Mn),
\label{static<S^x_piS^x_pi>}
\end{eqnarray}
\end{mathletters}\noindent
where $c$'s are numerical constants.
For $M\ll1$ the parameter $\eta$ is given by\cite{Bogoliubov}
\begin{equation}
\eta=2-\frac{4}{3}M+{\cal O}(M^2).
\label{theta}
\end{equation}
In general $\eta<2$ for $0<M<1$, and $\eta\to2$ when
$M\to0,1$.\cite{Haldane80} 
Thus, the leading term in $\langle S^z_{0,n+1}S^z_{0,1}\rangle$, besides
the constant $M^2$, is $n^{-\eta}\cos(2\pi Mn)$, which reflects the fact
that the hard-core bosons $t_{i,+}$ have a tendency to form a density wave
with a period equal to $1/M$.
On the other hand, the correlation functions
$\langle S^z_{\pi,i}S^z_{\pi,j}\rangle$ and
$\langle S^x_{0,i}S^x_{0,j}\rangle$ involve the massive $t_{i,0}$ and
$t_{i,-}$ bosons, and therefore decay exponentially for $|i-j|\gg1$.
We conclude that the correlators $\langle S^z_0S^z_0\rangle$ and
$\langle S^\pm_\pi S^\mp_\pi\rangle$ show quasi-long-range order while
$\langle S^z_\pi S^z_\pi\rangle$ and $\langle S^\pm_0S^\mp_0\rangle$ are
short-ranged.
These correlation functions are discussed in more detail in the next
section. 

Before closing this section, we make a few comments.
First, Eqs.\ (\ref{static<S^z_0S^z_0>}) and (\ref{static<S^x_piS^x_pi>})
have the same form as the correlators $\langle S^z_{n+1}S^z_1\rangle$
and $\langle S^x_{n+1}S^x_1\rangle$ of the $S=1$ Heisenberg chain in a
magnetic field larger than the Haldane
gap.\cite{Tsvelik,AffleckBose,Sakai} 
An important difference is that $\eta\ge2$ in the $S=1$
chain.\cite{Sakai} 
We will come back to this point in the next section.
Second, from the knowledge of the exact propagator of hard-core
bosons,\cite{Vaidya,Jimbo} we may expect that the correlation function
$\langle S^x_{\pi,n+1}S^x_{\pi,1}\rangle$ should have terms
proportional to $\cos(2\pi lMn)$ with $l=0,1,2,\ldots$.
Third, as noted in Introduction and will be discussed in Sec.\ IV,
the $t_{i,+}$ boson and $t_{i,0}$ boson correspond to a pair of holes
sitting on a rung and to a magnon in the SO(5) symmetric ladder,
respectively.
Thus, we may compare the hole-pair correlation and the spin
correlation functions with $\langle S^x_{\pi,i}S^x_{\pi,j}\rangle$
and $\langle S^z_{\pi,i}S^z_{\pi,j}\rangle$ of our model.

\section{Weak-coupling approach}

In this section we calculate the dynamical spin-spin correlation functions
in the weak-coupling limit, $J_\perp\ll J$.
We use the Abelian bosonization method which has been successfully applied
to the Heisenberg ladder.\cite{Schulz86,Strong,Shelton,Chitra}
In this approach we first bosonize two independent spin-$\frac{1}{2}$
Heisenberg chains and then treat the interchain coupling $J_\perp$
perturbatively. 
It is a relevant perturbation and is renormalized to a
strong-coupling regime, generating a mass gap in the excitation spectrum.
This behavior does not depend on the sign of $J_\perp$, and therefore the
model describes both the antiferromagnetic Heisenberg ladder for
$J_\perp>0$ and the $S=1$ Heisenberg chain for $J_\perp<0$.\cite{Shelton}
The spin gap in the latter case is nothing but the Haldane
gap.\cite{Haldane}
Since the gapless phase in a magnetic field is a single phase for
$0<J_\perp<\infty$, we expect the correlation functions in the
weak-coupling limit to have the same structure as in the
strong-coupling limit. 

We follow the formulation of Shelton {\it et al.},\cite{Shelton} which we
explain below to establish the notation.
We begin with the bosonized Hamiltonian of the ladder in the continuum
limit. 
It consists of three parts:
${\cal H}={\cal H}_++{\cal H}_-+{\cal H}_\perp$, where
\begin{eqnarray}
{\cal H}_+&=&\int dx\left\{
\frac{v}{2}\left[
 \left(\frac{d\phi_+}{dx}\right)^2
 +\left(\frac{d\theta_+}{dx}\right)^2
\right]
-\frac{m}{\pi a_0}\cos(\sqrt{4\pi}\phi_+)
-\frac{H}{\sqrt{\pi}}\frac{d\phi_+}{dx}\right\},
\label{H_+}\\
{\cal H}_-&=&\int dx\left\{
\frac{v}{2}\left[
 \left(\frac{d\phi_-}{dx}\right)^2
 +\left(\frac{d\theta_-}{dx}\right)^2
\right]
+\frac{2m}{\pi a_0}\cos(\sqrt{4\pi}\theta_-)
+\frac{m}{\pi a_0}\cos(\sqrt{4\pi}\phi_-)
\right\},
\label{H_-}\\
{\cal H}_\perp&=&
\frac{J_\perp}{2\pi^2a_0}\int dx\cos(\sqrt{4\pi}\theta_-)
\left[\cos(\sqrt{4\pi}\phi_+)-\cos(\sqrt{4\pi}\phi_-)\right]
+\frac{J_\perp a_0}{4\pi}\int dx
\left[\left(\frac{d\phi_+}{dx}\right)^2
     -\left(\frac{d\phi_-}{dx}\right)^2\right].
\label{H_perp}
\end{eqnarray}
Here $v$ is the spin-wave velocity, $a_0$ is a lattice constant, and
$m=J_\perp\lambda^2/2\pi$ with $\lambda$ being a numerical constant.
The bosonic fields $\phi_\pm$ and $\theta_\pm$ obey the commutation
relations
$[\phi_+(x),\theta_+(y)]=[\phi_-(x),\theta_-(y)]=i\Theta(y-x)$ and
$[\phi_+(x),\phi_-(y)]=[\phi_+(x),\theta_-(y)]
 =[\phi_-(x),\theta_+(y)]=[\theta_+(x),\theta_-(y)]=0$, where
$\Theta(x)$ is the step function.
It is important to note that $H$ appears in ${\cal H}_+$ only.
Thus, the external magnetic field changes the dynamics of the
fields $\phi_+$ and $\theta_+$, while the other fields $\phi_-$ and
$\theta_-$ are not directly influenced by the uniform field $H$.
The excitations involving these latter fields remain gapful even
above the lower critical field $H_{c1}$.
The spin operators with $q_y=0$ and $\pi$, defined in
Eqs.\ (\ref{S^z_0,i})--(\ref{S^-_pi,i}), are written in terms of the
bosonic fields as
\begin{mathletters}
\begin{eqnarray}
S^z_0(x)&=&
\frac{a_0}{\sqrt{\pi}}\frac{d\phi_+}{dx}
-(-1)^{x/a_0}\frac{2\lambda}{\pi}
    \sin(\sqrt{\pi}\phi_+)\cos(\sqrt{\pi}\phi_-),
\label{S^z_0(x)}\\
S^z_\pi(x)&=&
\frac{a_0}{\sqrt{\pi}}\frac{d\phi_-}{dx}
-(-1)^{x/a_0}\frac{2\lambda}{\pi}
    \cos(\sqrt{\pi}\phi_+)\sin(\sqrt{\pi}\phi_-),
\label{S^z_pi(x)}\\
S^+_0(x)&=&
\frac{2}{\pi}e^{i\sqrt{\pi}\theta_+}
\left[(-1)^{x/a_0}\lambda\cos(\sqrt{\pi}\theta_-)
+\cos(\sqrt{\pi}\theta_-)\sin(\sqrt{\pi}\phi_+)\cos(\sqrt{\pi}\phi_-)
+i\sin(\sqrt{\pi}\theta_-)\cos(\sqrt{\pi}\phi_+)\sin(\sqrt{\pi}\phi_-)
\right],
\nonumber\\&&
\label{S^+_0(x)}\\
S^+_\pi(x)&=&
\frac{2}{\pi}e^{i\sqrt{\pi}\theta_+}
\left[i(-1)^{x/a_0}\lambda\sin(\sqrt{\pi}\theta_-)
+\cos(\sqrt{\pi}\theta_-)\cos(\sqrt{\pi}\phi_+)\sin(\sqrt{\pi}\phi_-)
+i\sin(\sqrt{\pi}\theta_-)\sin(\sqrt{\pi}\phi_+)\cos(\sqrt{\pi}\phi_-)
\right],
\nonumber\\&&
\label{S^+_pi(x)}
\end{eqnarray}
\end{mathletters}\noindent
where $S^\alpha_{q_y}(x)=S^\alpha_{q_y,i}$ for $x=ia_0$
($\alpha=x,y,z$). 
Equations (\ref{S^z_0(x)})-(\ref{S^+_pi(x)}) directly follow from
Eq.~(30) and Appendix A of Ref.~\onlinecite{Shelton}.

Shelton {\it et al}.\ showed that ${\cal H}_+$ and ${\cal H}_-$ can be
greatly simplified by fermionization.
The Hamiltonian ${\cal H}_-$ then becomes that of free massive
Majorana fermions $\xi_3(x)$ and $\rho(x)$ having the mass gap $m$ and
$-3m$, respectively.
This is equivalent to the two-dimensional Ising model above or below the
critical temperature.
This observation allowed them to obtain the dynamical spin-spin
correlation functions\cite{Greven} from the known results of the Ising
model.\cite{Wu}
Physically the $\xi_3$ fermion describes the $S^z=0$ magnon
excitation, whereas the $\rho$ corresponds to a singlet excitation
with much higher energy.

On the other hand, the $\phi_+$ and $\theta_+$ fields represent the
$S^z=\pm1$ magnon excitations.
As the $S^z=1$ bosons condense above $H_{c1}$, these bosonic fields
have massless excitations.
We will first integrate out the massive Majorana fermions and
concentrate on the massless bosonic fields.
To proceed, here we introduce an approximation for dealing with
${\cal H}_\perp$. 
This interaction Hamiltonian has three components.
The first component involving only $\phi_-$ and $\theta_-$, i.e.,
$\cos(\sqrt{4\pi}\theta_-)\cos(\sqrt{4\pi}\phi_-)$ and
$(d\phi_-/dx)^2$, has two major effects on the dynamics of
${\cal H}_-$.
One effect is to renormalize the bare mass $m$ to $m\ln(\Lambda/m)$,
where $\Lambda$ is a high-energy cutoff, as noted by Shelton {\it et
  al.}\cite{Shelton}
This can be absorbed by redefining the mass. 
The other effect is a strong two-particle collision described by a
${\cal S}$ matrix having a superuniversal form, as recently discussed
by Damle and Sachdev.\cite{Damle}
Since we are only concerned with processes in which at most one
$S^z=0$ magnon is created, this strong scattering effect may be
irrelevant for our discussion of the dynamical correlations at zero
temperature.
The second component is a coupling term,
$\cos(\sqrt{4\pi}\theta_-)\cos(\sqrt{4\pi}\phi_+)$.
When integrating out the $\theta_-$ field perturbatively, we find that
the leading term
$\langle\cos(\sqrt{4\pi}\theta_-)\rangle_-\cos(\sqrt{4\pi}\phi_+)
 \propto\ln(\Lambda/m)\cos(\sqrt{4\pi}\phi_+)$
gives the renormalization of the mass of the $\phi_+$ and $\theta_+$
fields, $m\to m\ln(\Lambda/m)$, as expected from the SO(3) symmetry,
where the average is taken in the ground state of ${\cal H}_-$.
This is again taken care of by redefining the mass.
The higher-order terms will generate, through gradient expansions,
both irrelevant terms like
$\cos(\sqrt{4\pi}l\phi_+)$ with $l>1$, which we can safely ignore, and
a marginal operator $(\partial_x\phi_+)^2$, which should be kept.
The third component is the term $(d\phi_+/dx)^2$ already present in
${\cal H}_\perp$ and will be kept in the following calculation.
Hence we reduce ${\cal H}_\perp$ to the form
\begin{equation}
{\cal H}_\perp\approx
\frac{J_\perp a_0}{4\pi}\int dx
\left(\frac{d\phi_+}{dx}\right)^2,
\label{H_perp2}
\end{equation}
where the coupling constant $J_\perp a_0$ may be modified from its
bare value because of the emergence of the operator
$(\partial_x\phi_+)^2$ in the higher-order terms discussed above.
We believe, however, that the sign of the coupling constant is not
changed by the renormalization.\cite{Usami}
Having made this approximation, we now integrate out the
$\phi_-$ and $\theta_-$ fields to get the spin-spin correlation
functions. 
Within our approximation the fields $\phi_-$ and $\theta_-$ are
independent of $\phi_+$ and $\theta_+$, and therefore the correlation
functions of $\phi_-$ and $\theta_-$ are independent of $H$.
We use Eq.\ (33) of Ref.\ \onlinecite{Shelton} to represent
$\cos(\sqrt{\pi}\phi_-)$, $\sin(\sqrt{\pi}\phi_-)$,
$\cos(\sqrt{\pi}\theta_-)$, and $\sin(\sqrt{\pi}\theta_-)$ in terms of
the order and disorder parameters of an Ising model.
We then use Eqs.\ (38) and (39) of Ref.\ \onlinecite{Shelton} to
obtain their correlation functions.
The correlation functions $\langle S^\pm_0S^\mp_0\rangle$ involve
$\cos(\sqrt{\pi}\phi_-)\cos(\sqrt{\pi}\theta_-)$
and $\sin(\sqrt{\pi}\phi_-)\sin(\sqrt{\pi}\theta_-)$.
They are equivalent to free massive Majorana fermions, $\xi_3$, whose
correlators are easily obtained.
Finally, the correlator
$\langle\partial_x\phi_-(x)\partial_x\phi_-(0)\rangle$
decays exponentially [$\propto e^{-2mr/v}$ with
$r=(x^2+v^2\tau^2)^{1/2}$] and is ignored.
Hence, we arrive at the following expression of
the dynamical spin-spin correlation functions:
\begin{mathletters}
\begin{eqnarray}
\langle S^z_0(x,\tau)S^z_0(0,0)\rangle&=&
\frac{a_0^2}{\pi}
\langle\partial_x\phi_+(x,\tau)\partial_x\phi_+(0,0)\rangle_+,
\label{<S^z_0S^z_0>}\\
\langle S^z_\pi(x,\tau)S^z_\pi(0,0)\rangle&=&
(-1)^{x/a_0}\left(\frac{2\lambda}{\pi}\right)^2\frac{A^2_1}{\pi}
K_0(mr/v)
\langle
\cos[\sqrt{\pi}\phi_+(x,\tau)]\cos[\sqrt{\pi}\phi_+(0,0)]
\rangle_+,
\label{<S^z_piS^z_pi>}\\
\langle S^+_\pi(x,\tau)S^-_\pi(0,0)\rangle&=&
(-1)^{x/a_0}\left(\frac{2\lambda}{\pi}\right)^2A^2_1
\left\langle
e^{i\sqrt{\pi}\theta_+(x,\tau)}e^{-i\sqrt{\pi}\theta_+(0,0)}
\right\rangle_+,
\label{<S^+_piS^-_pi>}\\
\langle S^+_0(x,\tau)S^-_0(0,0)\rangle&=&
-\frac{a_0}{2\pi^2v}
\left\langle
e^{i\sqrt{\pi}\theta_+(x,\tau)}e^{-i\sqrt{\pi}\phi_+(x,\tau)}
e^{i\sqrt{\pi}\phi_+(0,0)}e^{-i\sqrt{\pi}\theta_+(0,0)}
\right\rangle_+
\partial_+K_0(mr/v)\nonumber\\
&&
-\frac{a_0}{2\pi^2v}
\left\langle
e^{i\sqrt{\pi}\theta_+(x,\tau)}e^{i\sqrt{\pi}\phi_+(x,\tau)}
e^{-i\sqrt{\pi}\phi_+(0,0)}e^{-i\sqrt{\pi}\theta_+(0,0)}
\right\rangle_+
\partial_-K_0(mr/v)\nonumber\\
&&
+\frac{a_0m}{2\pi^2v}K_0(mr/v)
\left[
\left\langle
e^{i\sqrt{\pi}\theta_+(x,\tau)}e^{-i\sqrt{\pi}\phi_+(x,\tau)}
e^{-i\sqrt{\pi}\phi_+(0,0)}e^{-i\sqrt{\pi}\theta_+(0,0)}
\right\rangle_+
\right.\nonumber\\
&&\hspace*{2.7cm}\left.
+\left\langle
e^{i\sqrt{\pi}\theta_+(x,\tau)}e^{i\sqrt{\pi}\phi_+(x,\tau)}
e^{i\sqrt{\pi}\phi_+(0,0)}e^{-i\sqrt{\pi}\theta_+(0,0)}
\right\rangle_+
\right],
\label{<S^+_0S^-_0>}
\end{eqnarray}
\end{mathletters}\noindent
where $\tau$ is the imaginary time,
$A_1$ is a numerical constant,
$\partial_\pm=\partial_\tau\pm iv\partial_x$, and
$\langle\ \rangle_+$ is the average with respect to the Hamiltonian
${\cal H}_+ + {\cal H}_\perp$.
In Eqs.\ (\ref{<S^z_0S^z_0>})--(\ref{<S^+_0S^-_0>}) we have ignored the
terms decaying much faster than $e^{-mr/v}$.
Therefore we discarded the contribution from processes involving more
than one massive magnons.
These equations are valid for $J_\perp>0$.
When $J_\perp<0$, on the other hand, the strongly renormalized interchain
coupling combines the spins $\mbox{\boldmath$S$}_{1,i}$ and
$\mbox{\boldmath$S$}_{2,i}$ into a single spin, and the ladder behaves as
a $S=1$ Heisenberg chain.
As explained in Ref.\ \onlinecite{Shelton}, when taking average over
the massive Majorana fermions $\xi_3$ and $\rho$, we only need to
exchange the order and disorder parameter of the Ising model.
Using the correlators of $\phi_-$ and $\theta_-$ for $J_\perp<0$ in
Ref.~\onlinecite{Shelton}, we find
\begin{mathletters}
\begin{eqnarray}
\langle S^z_0(x,\tau)S^z_0(0,0)\rangle\bigr|_{J_\perp<0}&=&
\langle S^z_0(x,\tau)S^z_0(0,0)\rangle\bigr|_{J_\perp>0}
+\langle S^z_\pi(x,\tau)S^z_\pi(0,0)\rangle\bigr|_{J_\perp>0},
\label{S^zS^z|J<0}\\
\langle S^+_0(x,\tau)S^-_0(0,0)\rangle\bigr|_{J_\perp<0}&=&
\langle S^+_0(x,\tau)S^-_0(0,0)\rangle\bigr|_{J_\perp>0}
+\langle S^+_\pi(x,\tau)S^-_\pi(0,0)\rangle\bigr|_{J_\perp>0},
\label{S^+S^-|J<0}
\end{eqnarray}
\end{mathletters}\noindent
where the correlators in the right-hand side are those in
Eqs.\ (\ref{<S^z_0S^z_0>})--(\ref{<S^+_0S^-_0>}).
When $J_\perp<0$, the $q_y=\pi$ correlators decay much faster than
the $q_y=0$ correlator and are thus negligible. 
Hence, the dynamical spin-spin correlation functions of the $S=1$ Haldane
chain are linear combinations of those $q_y=0$ and $\pi$ correlators of
the Heisenberg ladder ($J_\perp>0$).

Now our task is to calculate the correlators of $\phi_+$ and $\theta_+$
in the presence of the magnetic field.
The Hamiltonian ${\cal H}_+$ is identical to the one used to study the
commensurate-incommensurate transition in classical two-dimensional
systems.\cite{Pokrovsky,Okwamoto,Schulz80,Haldane82} 
In fact some of the correlation functions in
Eqs.\ (\ref{<S^z_0S^z_0>})--(\ref{<S^+_0S^-_0>}) have been discussed in
this context.\cite{Pokrovsky,Okwamoto,Schulz80,Haldane82}
In particular, the leading term of the correlation function corresponding
to
$\langle e^{i\sqrt{\pi}\phi_+(x,\tau)}
 e^{-i\sqrt{\pi}\phi_+(0,0)}\rangle$ is obtained by Schulz\cite{Schulz80}
including its universal exponent in the limit $M\to0$.
These results are used to calculate the spin-spin correlation
functions by Chitra and Giamarchi,\cite{Chitra} who unfortunately seem
to have overlooked some terms including the leading term
[$\propto\cos(2\pi Mx)$]
in $\langle S^z_0(x)S^z_0(0)\rangle$.
We think therefore that it is still worthwhile to describe the
calculation of the correlations of $\phi_+$ and $\theta_+$ in
Eqs.\ (\ref{<S^z_0S^z_0>})--(\ref{<S^+_0S^-_0>}) in some detail,
despite the fact that the Hamiltonian ${\cal H}_+$ has been analyzed
in many literatures. 
To our knowledge, our results concerning the massive modes are new.

\begin{figure}
\begin{center}
\leavevmode\epsfxsize=65mm
\epsfbox{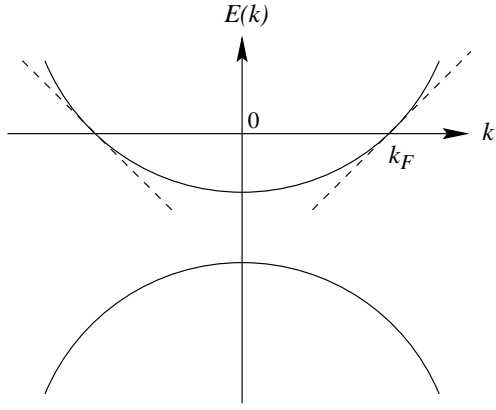}
\end{center}
\caption{Schematic picture of the upper and lower bands:
$E(k)=\pm{\protect\sqrt{v^2k^2+m^2}}-H$.
The negative-energy states are filled.
The long-distance behavior of the correlation functions are determined by
the low-energy excitations around $|k|=k_F$, where the dispersion is
linearized. 
}
\label{fig:band}
\end{figure}

Following Ref.\ \onlinecite{Shelton}, we fermionize ${\cal H}_+$:
\begin{eqnarray}
{\cal H}_+&=&
\int dx\left[
iv\left(\psi^\dagger_L\frac{d}{dx}\psi_L
        -\psi^\dagger_R\frac{d}{dx}\psi_R\right)
-im\left(\psi^\dagger_R\psi_L-\psi^\dagger_L\psi_R\right)
-H\left(\psi^\dagger_L\psi_L+\psi^\dagger_R\psi_R\right)
\right],
\nonumber\\
&=&
\int^\infty_{-\infty}dk
\left[
vk\left(c^\dagger_{R,k}c_{R,k}-c^\dagger_{L,k}c_{L,k}\right)
-im\left(c^\dagger_{R,k}c_{L,k}-c^\dagger_{L,k}c_{R,k}\right)
-H\left(c^\dagger_{R,k}c_{R,k}+c^\dagger_{L,k}c_{L,k}\right)
\right],
\label{H_+2}
\end{eqnarray}
where $\psi_L$ ($\psi_R$) is the right-going (left-going) complex fermion
field,  and $\psi_{R(L)}(x)=\int(dk/\sqrt{2\pi})e^{ikx}c_{R(L),k}$.
The fermion fields are related to the bosons by the standard relations:
\begin{mathletters}
\begin{eqnarray}
\psi_R(x)&=&
\frac{1}{\sqrt{2\pi a_0}}e^{i\sqrt{\pi}[\phi_+(x)-\theta_+(x)]},\quad
:\!\psi^\dagger_R(x)\psi_R(x)\!:\,=
\frac{1}{2\sqrt{\pi}}\frac{d}{dx}[\phi_+(x)-\theta_+(x)],
\label{psi_R}\\
\psi_L(x)&=&
\frac{1}{\sqrt{2\pi a_0}}e^{-i\sqrt{\pi}[\phi_+(x)+\theta_+(x)]},\quad
:\!\psi^\dagger_L(x)\psi_L(x)\!:\,=
\frac{1}{2\sqrt{\pi}}\frac{d}{dx}[\phi_+(x)+\theta_+(x)].
\label{psi_L}
\end{eqnarray}
\end{mathletters}\noindent
It is important to note that the normal ordering in the above equations is
defined with respect to the ground state of $H=0$.
The fermionized Hamiltonian ${\cal H}_+$ is easily diagonalized:
\begin{equation}
{\cal H}_+=
\int^\infty_{-\infty}dk
\left[
\left(\sqrt{v^2k^2+m^2}-H\right)a^\dagger_ka_k
-\left(\sqrt{v^2k^2+m^2}+H\right)\tilde a^\dagger_k\tilde a_k
\right],
\label{diagonalH_+}
\end{equation}
where
\begin{equation}
\pmatrix{a_k\cr \tilde a_k\cr}=
\pmatrix{
\cos(\varphi_k/2) & -i\sin(\varphi_k/2) \cr
-i\sin(\varphi_k/2) & \cos(\varphi_k/2) \cr}
\pmatrix{c_{R,k} \cr c_{L,k} \cr}
\label{unitary}
\end{equation}
with $\tan\varphi_k=m/vk$.
The magnetic field plays a role of the chemical potential to the fermions.
We are concerned with the case where $H$ is slightly above the lower
critical field $H_{c1}(=m)$ such that $0<M\ll1$.
The ground state is obtained by filling the upper band ($a_k$) up to
the Fermi points ($|k|<k_F$) and the lower band ($\tilde a_k$) completely;
see Fig.~\ref{fig:band}. 
The Fermi wavenumber $k_F$ is related to the magnetization:
$k_F=\pi M/a_0$. 
This follows from
\begin{equation}
\frac{M}{a_0}=
\frac{1}{\sqrt{\pi}L}\int dx\langle\partial_x\phi_+\rangle_+
=\frac{1}{L}\int dx
\langle\,:\!\psi^\dagger_R\psi_R+\psi^\dagger_L\psi_L\!:\,\rangle_+
=\frac{1}{2\pi}\int^\infty_{-\infty}dk\langle a^\dagger_ka_k\rangle_+,
\label{k_F}
\end{equation}
where $L$ is the length of the ladder and
$\langle a^\dagger_ka_k\rangle_+=\Theta(k_F-|k|)$.
In calculating long-distance correlations, we can safely ignore the
lower band and keep only the low-energy excitations around the Fermi
points in the upper band. 
In the fermion representation the interaction term ${\cal H}_\perp$
reads
\begin{equation}
{\cal H}_\perp\approx
\frac{J_\perp a_0}{4}\int dx\left(\psi^\dagger\psi\right)^2,
\label{Hint}
\end{equation}
where $\psi(x)=\int(dk/\sqrt{2\pi})e^{ikx}a_k$ and we have dropped the
contribution from the lower band.
The total Hamiltonian for the fermions in the upper band,
${\cal H}_a={\cal H}_++{\cal H}_\perp$, consists of the kinetic
energy, Eq.\ (\ref{diagonalH_+}), and the short-range scattering term,
Eq.\ (\ref{Hint}).
The coupling constant of the latter term is proportional to $J_\perp$
in lowest order.
Thus, the interaction is repulsive for the antiferromagnetically
coupled ladder, while it is attractive for the $S=1$
chain.\cite{Usami} 
Obviously the scattering term has only negligible effects in both
limits $M\to0$ and $M\to1$, where we will get the correlation
functions of free fermions.\cite{Haldane80}

The low-energy physics of ${\cal H}_a$ can be easily solved by the
Abelian bosonization.\cite{Emery,Solyom}
We first linearize the dispersion around $k=\pm k_F$
(Fig.~\ref{fig:band}). 
We then bosonize the fermions in the upper band:
\begin{equation}
\psi(x)
\approx
\frac{1}{\sqrt{2\pi a_0}}\left(
e^{i\pi Mx/a_0+i\sqrt{\pi}[\phi(x)-\theta(x)]}
+e^{-i\pi Mx/a_0-i\sqrt{\pi}[\phi(x)+\theta(x)]}
\right),
\label{bosonize}
\end{equation}
where the bosonic fields $\phi(x)$ and $\theta(x)$ obey
$[\phi(x),\theta(y)]=i\Theta(y-x)$.
Using these fields, we write the Hamiltonian ${\cal H}_a$ as
\begin{equation}
{\cal H}_a=
\frac{\tilde v}{2}\int dx\left[
\frac{1}{g}\left(\frac{d\phi}{dx}\right)^2
+g\left(\frac{d\theta}{dx}\right)^2
\right],
\label{H_+boson}
\end{equation}
where $\tilde v$ is the Fermi velocity, and $g$ is a parameter
determined by the interaction: $g<1$ ($g>1$) when $J_\perp>0$
($J_\perp<0$), and $g\to1$ as $M\to0,1$.
Incidentally, $g$ is related to the compactification radius $R$ of the
field $\phi$ by $g=1/(4\pi R^2)$.
We now need to express $\phi_+$ and $\theta_+$ in terms of $\phi$ and
$\theta$.
Once this is done, it is straightforward to calculate the correlation
functions since ${\cal H}_a$ is a free-boson Hamiltonian.
First we note that for states near the Fermi surface we have
\begin{equation}
c^\dagger_{R,k}c_{R,p}+c^\dagger_{L,k}c_{L,p}\approx
a^\dagger_ka_p+\tilde a^\dagger_k\tilde a_p,
\label{approximation}
\end{equation}
where we used the approximation
$\varphi_k\approx\varphi_p\approx\varphi_{k_F}$.
Using Eqs.\ (\ref{bosonize}) and (\ref{approximation}) and discarding the
$\tilde a_k$ fermions, we find
\begin{equation}
\frac{1}{\sqrt{\pi}}\frac{d\phi_+}{dx}=\,
:\!\psi^\dagger_R(x)\psi_R(x)+\psi^\dagger_L(x)\psi_L(x)\!:\,
\approx\psi^\dagger(x)\psi(x)
=\frac{M}{a_0}+\frac{1}{\sqrt{\pi}}\frac{d\phi}{dx}
+\frac{1}{\pi a_0}\cos[2\pi Mx+\sqrt{4\pi}\phi(x)].
\label{dphi_+/dx}
\end{equation}
It follows that
\begin{eqnarray}
\frac{1}{\pi}
\langle\partial_x\phi_+(x,\tau)\partial_x\phi_+(0,0)\rangle_+
&=&
\frac{M^2}{a_0^2}
+\frac{1}{\pi}
\langle\partial_x\phi(x,\tau)\partial_x\phi(0,0)\rangle_a
\nonumber\\
&&
+\frac{1}{(\pi a_0)^2}\cos(2\pi Mx)
\langle\cos[\sqrt{4\pi}\phi(x,\tau)]\cos[\sqrt{4\pi}\phi(0,0)]\rangle_a,
\label{<dphi_+dphi_+>}
\end{eqnarray}
where $\langle\ \rangle_a$ represents the average taken in the ground
state of ${\cal H}_a$.
The averages are found to be
\begin{eqnarray}
&&
\langle\partial_x\phi(x,\tau)\partial_x\phi(0,0)\rangle_a=
-\frac{g}{4\pi}\left(
 \frac{1}{(x+i\tilde v\tau)^2}+\frac{1}{(x-i\tilde v\tau)^2}\right),
\label{<dphidphi>a}\\
&&
\langle\cos[\sqrt{4\pi}\phi(x,\tau)]\cos[\sqrt{4\pi}\phi(0,0)]\rangle_a=
\frac{1}{2}
\left(\frac{\tilde a_0^2}{x^2+\tilde v^2\tau^2}\right)^g,
\label{<cos(4piphi)cos(4piphi)>a}
\end{eqnarray}
where $\tilde a_0$ is a short-distance cutoff of order $a_0$. 
Integration of Eq.\ (\ref{dphi_+/dx}) then yields
\begin{equation}
\phi_+(x,\tau)=\frac{\sqrt{\pi}Mx}{a_0}+\phi(x,\tau).
\label{phi_+}
\end{equation}
We have neglected the contribution from the oscillating term.
We thus get
\begin{eqnarray}
\langle\cos[\sqrt{\pi}\phi_+(x,\tau)]\cos[\sqrt{\pi}\phi_+(0,0)]\rangle_+
&=&\cos(\pi Mx/a_0)
\langle\cos[\sqrt{\pi}\phi(x,\tau)]\cos[\sqrt{\pi}\phi(0,0)]\rangle_a
\nonumber\\
&=&
\frac{1}{2}\cos(\pi Mx/a_0)
\left(\frac{\tilde a_0^2}{x^2+\tilde v^2\tau^2}\right)^{g/4}.
\label{<cos(piphi)cos(piphi)>a}
\end{eqnarray}
We next consider $e^{-i\sqrt{\pi}\theta_+(x)}$.
From Eqs.\ (\ref{psi_R}) and (\ref{psi_L}) we can express it as
\begin{equation}
e^{-i\sqrt{\pi}\theta_+(x)}=
\sqrt{\frac{\pi a_0}{2}}\left[
e^{i\pi/4}e^{-i\sqrt{\pi}\phi_+(x)}\psi_R(x)
+e^{-i\pi/4}e^{i\sqrt{\pi}\phi_+(x)}\psi_L(x)\right].
\label{e^(-ipitheta)}
\end{equation}
Using the same approximation as in the derivation of
Eqs.\ (\ref{approximation}) and (\ref{dphi_+/dx}), we get
$\psi_R(x)\approx\psi(x)/\sqrt{2}$ and
$\psi_L(x)\approx i\psi(x)/\sqrt{2}$.
Here we have made a further approximation $\varphi_{k_F}\approx\pi/2$
which is valid for $0<M\ll1$.
From Eqs.\ (\ref{bosonize}), (\ref{phi_+}), and (\ref{e^(-ipitheta)}) we
find 
\begin{equation}
e^{-i\sqrt{\pi}\theta_+(x)}=
\left\{
\frac{1}{\sqrt{2}}+\sin[2\pi Mx/a_0+\sqrt{4\pi}\phi(x)+(\pi/4)]
\right\}
e^{-i\sqrt{\pi}\theta(x)+i\pi/4}.
\label{e^(-ipitheta)-2}
\end{equation}
From Eq.\ (\ref{e^(-ipitheta)-2}) we may write the correlation
functions in Eqs.\ (\ref{<S^+_piS^-_pi>}) and (\ref{<S^+_0S^-_0>}) as
\begin{eqnarray*}
\left\langle e^{i\sqrt{\pi}\theta_+(x,\tau)}
   e^{-i\sqrt{\pi}\theta_+(0,0)}\right\rangle_+
&=&
\frac{1}{2}
\left\langle
e^{i\sqrt{\pi}\theta(x,\tau)}e^{-i\sqrt{\pi}\theta(0,0)}
\right\rangle_a
\\
&&
+\frac{1}{4}\sum_{\epsilon=\pm1}e^{2\pi i\epsilon Mx/a_0}
\left\langle
 e^{i\sqrt{\pi}\theta(x,\tau)}e^{i\epsilon\sqrt{4\pi}\phi(x,\tau)}
 e^{-i\epsilon\sqrt{4\pi}\phi(0,0)}e^{-i\sqrt{\pi}\theta(0,0)}
\right\rangle_a,
\end{eqnarray*}
\begin{eqnarray*}
\left\langle
 e^{i\sqrt{\pi}\theta_+(x,\tau)}e^{-i\sqrt{\pi}\phi_+(x,\tau)}
 e^{i\sqrt{\pi}\phi_+(0,0)}e^{-i\sqrt{\pi}\theta_+(0,0)}
\right\rangle_+ \!\!
&=&
\frac{e^{-i\pi Mx/a_0}}{2}
\left\langle
 e^{i\sqrt{\pi}\theta(x,\tau)}e^{-i\sqrt{\pi}\phi(x,\tau)}
 e^{i\sqrt{\pi}\phi(0,0)}e^{-i\sqrt{\pi}\theta(0,0)}
\right\rangle_a
\\
&&
+\frac{e^{i\pi Mx/a_0}}{4}
\left\langle
 e^{i\sqrt{\pi}\theta(x,\tau)}e^{i\sqrt{\pi}\phi(x,\tau)}
 e^{-i\sqrt{\pi}\phi(0,0)}e^{-i\sqrt{\pi}\theta(0,0)}
\right\rangle_a
\\
&&
+\frac{e^{-3\pi iMx/a_0}}{4}
\left\langle
 e^{i\sqrt{\pi}\theta(x,\tau)}e^{-i\sqrt{9\pi}\phi(x,\tau)}
 e^{i\sqrt{9\pi}\phi(0,0)}e^{-i\sqrt{\pi}\theta(0,0)}
\right\rangle_a,
\end{eqnarray*}
and
\begin{eqnarray*}
&&
\sum_{\epsilon=\pm1}
\left\langle
 e^{i\sqrt{\pi}\theta_+(x,\tau)}e^{i\epsilon\sqrt{\pi}\phi_+(x,\tau)}
 e^{i\epsilon\sqrt{\pi}\phi_+(0,0)}e^{-i\sqrt{\pi}\theta_+(0,0)}
\right\rangle_+
\\
&&\qquad
=
\frac{1}{2}\sum_{\epsilon=\pm1}e^{i\epsilon\pi Mx/a_0}
\left\langle
 e^{i\sqrt{\pi}\theta(x,\tau)}e^{i\epsilon\sqrt{\pi}\phi(x,\tau)}
 e^{-i\epsilon\sqrt{\pi}\phi(0,0)}e^{-i\sqrt{\pi}\theta(0,0)}
\right\rangle_a.
\end{eqnarray*}
The averages in the above equations are given by
\begin{equation}
\langle e^{i\sqrt{\pi}\theta(x,\tau)}e^{in\sqrt{\pi}\phi(x,\tau)}
  e^{-in\sqrt{\pi}\phi(0,0)}e^{-i\sqrt{\pi}\theta(0,0)}\rangle_a=
e^{-i\pi n/2}
\left(\frac{\tilde a_0}{x+i\tilde v\tau}\right)
        ^{(1/\sqrt{g}-n\sqrt{g})^2/4}
\left(\frac{\tilde a_0}{x-i\tilde v\tau}\right)
        ^{(1/\sqrt{g}+n\sqrt{g})^2/4}.
\label{<e(itheta)e(inphi)e(-inphi)e(-itheta)>a}
\end{equation}
Combining these results together, we finally obtain the dynamical
spin-spin correlation functions:
\begin{eqnarray}
\langle S^z_0(x,\tau)S^z_0(0,0)\rangle&=&
M^2
-\frac{g}{4\pi^2}
 \left[\frac{1}{(x+i\tilde v\tau)^2}
       +\frac{1}{(x-i\tilde v\tau)^2}\right]
+C_1\cos(2\pi Mx)
\left(\frac{\tilde a_0^2}{x^2+\tilde v^2\tau^2}\right)^g,
\label{<S^z_0S^z_0>a}\\
\langle S^z_\pi(x,\tau)S^z_\pi(0,0)\rangle&=&
C_2(-1)^x\cos(\pi Mx)K_0(mr/v)
\left(\frac{\tilde a_0^2}{x^2+\tilde v^2\tau^2}\right)^{g/4},
\label{<S^z_piS^z_pi>a}\\
\langle S^+_\pi(x,\tau)S^-_\pi(0,0)\rangle&=&
(-1)^x
\left\{
C_3\left(\frac{\tilde a_0^2}{x^2+\tilde v^2\tau^2}\right)^{1/4g}
-C_4
\left(\frac{\tilde a_0^2}{x^2+\tilde v^2\tau^2}\right)
           ^{(1/2\sqrt{g}-\sqrt{g})^2}
\left[
\frac{\tilde a_0^2e^{2\pi iMx}}{(x+i\tilde v\tau)^2}
+\frac{\tilde a_0^2e^{-2\pi iMx}}{(x-i\tilde v\tau)^2}
\right]
\right\},
\label{<S^+_piS^-_pi>a}\\
\langle S^\pm_0(x,\tau)S^\mp_0(0,0)\rangle&=&
\frac{i}{8\pi^2v}
\left(\frac{\tilde a_0^2}{x^2+\tilde v^2\tau^2}\right)
^{(1/\sqrt{g}-\sqrt{g})^2/4}
\left[
\left(\frac{2\tilde a_0e^{\pm i\pi Mx}}{x-i\tilde v\tau}
  -\frac{\tilde a_0e^{\mp i\pi Mx}}{x+i\tilde v\tau}\right)
\partial_-K_0(mr/v)
+(x\to-x)\right]
\nonumber\\
&&
\pm\frac{im}{4\pi^2v}K_0(mr/v)
\left(\frac{\tilde a^2_0}{x^2+\tilde v^2\tau^2}\right)
^{(1/\sqrt{g}-\sqrt{g})^2/4}
\left(
\frac{\tilde a_0e^{\mp i\pi Mx}}{x+i\tilde v\tau}
-\frac{\tilde a_0e^{\pm i\pi Mx}}{x-i\tilde v\tau}
\right),
\label{<S^+_0S^-_0>a}
\end{eqnarray}
where $C$'s are positive numerical constants, and we have set $a_0=1$.
The correlation function
$\langle S^-_\pi(x,\tau)S^+_\pi(0,0)\rangle$ is obtained by replacing
$M$ with $-M$ in Eq.\ (\ref{<S^+_piS^-_pi>a}).
In Eq.\ (\ref{<S^+_0S^-_0>a})
we have discarded the term proportional to $e^{\pm 3\pi iMx}$
decaying much faster than the kept terms.
We should therefore regard
Eqs.\ (\ref{<S^z_0S^z_0>a})--(\ref{<S^+_0S^-_0>a}) as listing only the
leading terms.
As noted in Sec.\ II, we may expect that
$\langle S^\pm_\pi(x,\tau)S^\mp_\pi(0,0)\rangle$ should contain
algebraically decaying terms that are proportional to $\cos(2\pi lMx)$
with any integer $l$.
From Ref.\ \onlinecite{Haldane82}, we expect that the correlator
$\langle S^z_0(x)S^z_0(0)\rangle$ should also have terms proportional to
$\cos(2\pi lMx)$ which decay as $x^{-2l^2g}$.
The appearance of the term proportional to $e^{-3\pi iMx}$
suggests that $\langle S^\pm_0(x,\tau)S^\mp_0(0,0)\rangle$ should
have terms proportional to $K_0(mr/v)\cos[\pi(2l+1)Mx]$.
We note that the equal-time correlations
$\langle S^z_0(x,0)S^z_0(0,0)\rangle$ and
$\langle S^\pm_\pi(x,0)S^\mp_\pi(0,0)\rangle$ agree with
Eqs.\ (\ref{static<S^z_0S^z_0>}) and (\ref{static<S^x_piS^x_pi>}) if we
identify $\eta$ with $2g$.
As is well known, the strongest correlation is
$\langle S^\pm_\pi(x,0)S^\mp_\pi(0,0)\rangle\sim(-1)^xx^{-1/\eta}$.
Note also that the value of the exponent is consistent between the weak-
and strong-coupling approach: $g<1$ ($g>1$) for $J_\perp>0$ ($J_\perp<0$)
and $g\to1$ as $M\to0,1$.
Another interesting finding is that the exponentially decaying equal-time
correlation functions have different phases by $\pi/2$:
$\langle S^z_\pi(x,0)S^z_\pi(0,0)\rangle\propto\cos(\pi Mx)$ and
$\langle S^+_0(x,0)S^-_0(0,0)\rangle\propto\sin(\pi Mx)$.

Now we are in a position to calculate the dynamical spin structure factors
defined by
\begin{mathletters}
\begin{equation}
S^{\alpha\beta}_{q_y}(q,\omega)=
\frac{1}{2\pi}\int^\infty_{-\infty}dx\int^\infty_{-\infty}dt
\langle S^\alpha_{q_y}(x,\tau=it+0^+)S^\beta_{q_y}(0,0)\rangle
e^{-iqx+i\omega t},
\label{S(q,omega)-1}
\end{equation}
where $t$ is a real time and the correlation functions in the real time
are obtained by replacing $\tau\to it+0^+$.
We may also calculate it from
\begin{equation}
S^{\alpha\beta}_{q_y}(q,\omega)=
\frac{1}{\pi}{\rm Im}\lim_{i\tilde\omega\to\omega+i0^+}
\int^\infty_{-\infty}dx\int^\infty_{-\infty}d\tau
\langle S^\alpha_{q_y}(x,\tau)S^\beta_{q_y}(0,0)\rangle
e^{-iqx+i\tilde\omega\tau}
\label{S(q,omega)-2}
\end{equation}
\end{mathletters}\noindent
for $\omega>0$.
From the obvious relation
$S^{\alpha\beta}_{q_y}(q,\omega)=S^{\alpha\beta}_{q_y}(-q,\omega)
 =S^{\alpha\beta}_{q_y}(q+2\pi,\omega)$, we assume $0\le q\le\pi$ in the
following discussion.
We first consider the correlation functions showing the quasi-long-range
order.
Using Eq.\ (\ref{formula1}) in Appendix,
we get from Eqs.\ (\ref{<S^z_0S^z_0>a}) and (\ref{<S^+_piS^-_pi>a})
\begin{eqnarray}
S^{zz}_0(q,\omega)&=&
2\pi M^2\delta(q)\delta(\omega)
+\frac{g\omega}{2\pi\tilde v}\Theta(\omega)
 \delta(\omega-\tilde vq)
\nonumber\\
&&
+\frac{\pi C_1\tilde a_0^2}{2\tilde v[\Gamma(g)]^2}
\Theta\biglb(\omega-\tilde v(q-2\pi M)\bigrb)
\Theta\biglb(\omega+\tilde v(q-2\pi M)\bigrb)
\left(\frac{4\tilde v^2/\tilde a_0^2}
           {\omega^2-\tilde v^2(q-2\pi M)^2}\right)
  ^{1-g},
\label{S^zz_0(q,omega)}\\
S^{+-}_\pi(q,\omega)&=&
\frac{\pi C_3\tilde a_0^2}{\tilde v[\Gamma(1/4g)]^2}
\Theta\biglb(\omega+\tilde v(q-\pi)\bigrb)
\Theta\biglb(\omega-\tilde v(q-\pi)\bigrb)
\left(\frac{4\tilde v^2/\tilde a_0^2}
           {\omega^2-\tilde v^2(q-\pi)}\right)
  ^{1-1/4g}\nonumber\\
&&
+\frac{\pi C_4\tilde a_0^2}
      {\tilde v\Gamma(\eta_+)\Gamma(\eta_-)}
\biggl[
\Theta\biglb(\omega-\tilde v[q-\pi(1-2M)]\bigrb)
\Theta\biglb(\omega+\tilde v[q-\pi(1-2M)]\bigrb)
\nonumber\\
&&\hspace*{3.2cm}\times
\left(\frac{\omega-\tilde v[q-\pi(1-2M)]}
           {2\tilde v/\tilde a_0}\right)^{\eta_+-1}
\left(\frac{2\tilde v/\tilde a_0}
           {\omega+\tilde v[q-\pi(1-2M)]}\right)^{1-\eta_-}
\nonumber\\
&&\hspace*{2.5cm}
+\Theta\biglb(\omega-\tilde v[q-\pi(1+2M)]\bigrb)
 \Theta\biglb(\omega+\tilde v[q+\pi(1+2M)]\bigrb)
\nonumber\\
&&\hspace*{3.2cm}\times
\left(\frac{\omega+\tilde v[q-\pi(1+2M)]}
           {2\tilde v/\tilde a_0}\right)^{\eta_+-1}
\left(\frac{2\tilde v/\tilde a_0}
           {\omega-\tilde v[q-\pi(1+2M)]}\right)^{1-\eta_-}
\biggr],
\nonumber\\
&&
\label{S^+-_pi(q,omega)}
\end{eqnarray}
where $\eta_\pm=[(1/2\sqrt{g})\pm\sqrt{g}\,]^2$.
The structure factor $S^{-+}_\pi(q,\omega)$ can be obtained from
$S^{+-}_\pi(q,\omega)$ by $M\to-M$.
Being obtained from the long-distance asymptotic expansions
(\ref{<S^z_0S^z_0>a}) and (\ref{<S^+_piS^-_pi>a}), each term in
Eqs.\ (\ref{S^zz_0(q,omega)}) and (\ref{S^+-_pi(q,omega)}) describe
the behavior of the structure factors correctly only near its
low-energy threshold.
For example, the last term ($\propto C_1$) in
Eq.\ (\ref{S^zz_0(q,omega)}) is valid only for
$\omega\ll\pi\tilde vM$ ($|q-2\pi M|\ll\pi M$) and cannot be extended
to $|q|\lesssim\pi M$.
The supports of these structure factors are shown in
Figs.~\ref{fig:S^zz_0(q,omega)}--\ref{fig:S^-+_pi(q,omega)}.
They are essentially the same as those of the $S=\frac{1}{2}$ $XXZ$
chain\cite{Muller} except that the small $M$ in the ladder corresponds
to the nearly polarized state in the $S=\frac{1}{2}$ chain through the
relation $M=\frac{1}{2}+\langle\widetilde{S}^z\rangle$.
The strongest divergence is at $q=\pi$ of $S^{\pm\mp}_\pi(q,\omega)$:
$S^{\pm\mp}_\pi(q,\omega)\propto[\omega-\tilde v(q-\pi)]^{-1+1/4g}$.
The exponent approaches $-3/4$ as $M\to0,1$.
We note that the boundaries of the supports of these structure factors
became all straight lines because of our linearization of the dispersion
relation in the continuum limit.
This is an artifact of the approximation, and the true boundary lines
should be given by some nonlinear functions.
Furthermore, some of the boundary lines may be parts of a single
curve.

We next consider the massive components.
Using Eqs.\ (\ref{<S^z_piS^z_pi>a}), (\ref{S(q,omega)-2}), and
(\ref{formula2}), we get
\begin{eqnarray}
S^{zz}_\pi(q,\omega)&=&
C_2v\tilde v\left(\frac{\tilde a_0}{2\tilde v}\right)^{g/2}
\frac{\pi}{[\Gamma(g/4)]^2}\int^\infty_{-\infty}dk
\frac{\Theta\biglb(\omega-\varepsilon(q-k-\pi+\pi M)-\tilde v|k|\bigrb)}
  {\varepsilon(q-k-\pi+\pi M)
    \{[\omega-\varepsilon(q-k-\pi+\pi M)]^2-\tilde v^2k^2\}^{1-g/4}}
\nonumber\\
&&
+(M\to-M),
\label{S^zz_pi(q,omega)}
\end{eqnarray}
where $\varepsilon(q)=\sqrt{v^2q^2+m^2}$.
The minimum energy above which $S^{zz}_\pi(q,\omega)>0$ is
$\omega=\varepsilon\biglb(q-\pi(1-M)\bigrb)$ around $q=\pi(1-M)$.
Near this threshold energy the structure factor reduces to
\begin{equation}
S^{zz}_\pi(q,\omega)=
C_2\frac{\pi\tilde a_0v}{2m\tilde v\Gamma(g/2)}
\Theta\biglb(\omega-\varepsilon(q-\pi+\pi M)\bigrb)
\left(\frac{\tilde v/\tilde a_0}{\omega-\varepsilon(q-\pi+\pi M)}\right)
  ^{1-g/2}
+(M\to-M),
\label{S^zz_pi(q,omega)-2}
\end{equation}
where $v|q-\pi(1\mp M)|\ll m\tilde v/v$
and $|\omega-\varepsilon\biglb(q-\pi(1\mp M)\bigrb)|\ll m(\tilde
v/v)^2$ are assumed.
The support of $S^{zz}_\pi(q,\omega)$ is shown in
Fig.~\ref{fig:S^zz_pi(q,omega)}.
We see that $S^{zz}_\pi(q,\omega)$ diverges at the low-energy threshold
as
$S^{zz}_\pi(q,\omega)\propto[\omega-\varepsilon(q-\pi+\pi M)]^{-1+g/2}$.
Since the two thresholds $\omega=\varepsilon(q-\pi\pm\pi M)$
intersect at $q=\pi$, we expect to have a peak at $q=\pi$ and
$\omega=\varepsilon(\pi M)=H$.
The exponent approaches $-1/2$ as $M\to0,1$, and the singularity is even
stronger for $0<M<1$, where $g<1$.
Note that the exponent jumps from $-1$ to $-1/2$ when $H$ crosses $H_{c1}$
from below.
In the strong-coupling limit of the ladder, the square-root divergence
may be understood in the following way.
The correlation function $\langle S^z_{\pi,i}S^z_{\pi,j}\rangle$ is
a propagator of the $t_{i,0}$ bosons.
If we ignored the interaction with the $t_{i,+}$ bosons, we would get the
massive free-particle propagator, $K_0(mr/v)$.
Due to the interaction the motion of the $t_{i,0}$ boson is
necessarily  accompanied by a superfluid flow of the $t_{i,+}$ bosons.
Its main effect in the low-density limit amounts both to multiplying the
free propagator by that of the hard-core $t_{i,+}$ bosons
$\propto(x^2+\tilde v^2\tau^2)^{-1/4}$ and to shifting the momentum by
$\pi M$.
The Fourier transform of the product has the square-root divergence at the
threshold.

Finally we consider $S^{\pm\mp}_0(q,\omega)$.
This can be obtained from Eqs.\ (\ref{<S^+_0S^-_0>a}) and
(\ref{S(q,omega)-2}) as described in Appendix.
The result is
\begin{eqnarray}
S^{\pm\mp}_0(q,\omega)&=&
\frac{\tilde a_0}{4\pi\Gamma(\eta_0)\Gamma(\eta_0+1)}
\left(\frac{\tilde a_0}{2\tilde v}\right)^{2\eta_0}
\int^\infty_{-\infty}dk
\frac{\Theta\biglb(\omega-\varepsilon(q-k\mp\pi M)-\tilde v|k|\bigrb)
       [\omega-\varepsilon(q-k\mp\pi M)-\tilde vk]^{\eta_0}}
  {\varepsilon(q-k\mp\pi M)
    [\omega-\varepsilon(q-k\mp\pi M)+\tilde vk]^{1-\eta_0}}
\nonumber\\
&&\hspace*{5cm}
\times\left[
\frac{3}{2}\varepsilon(q-k\mp\pi M)
-\frac{1}{2}v(q-k\mp\pi M)\pm m\right]
\nonumber\\
&&
+(q\to-q),
\label{S^+-_0(q,omega)}
\end{eqnarray}
where $\eta_0=[(1/\sqrt{g})-\sqrt{g}\,]^2/4$.
Near the lower edge $0<\omega-\varepsilon(q\mp\pi M)\ll m$, it may be
approximated by
\begin{equation}
S^{+-}_0(q,\omega)=
\frac{5\tilde a_0}{8\pi\tilde v\Gamma(2\eta_0+1)}
\Theta\biglb(\omega-\varepsilon(q-\pi M)\bigrb)
\left(\frac{\omega-\varepsilon(q-\pi M)}{\tilde v/\tilde a_0}\right)
   ^{2\eta_0}
+(M\to-M).
\label{S^+-_0(q,omega)-2}
\end{equation}
In general the exponent $2\eta_0\ge0$ and approaches 0 as $M\to0,1$.
The support of $S^{+-}_0(q,\omega)$ is shown in
Fig.~\ref{fig:S^+-_0(q,omega)}.
Like $S^{zz}_\pi(q,\omega)$, $S^{+-}_0(q,\omega)$ has a peak at
$(q,\omega)=(0,H)$, where the two thresholds
$\omega=\varepsilon(q\pm\pi M)$ cross.
The structure factor $S^{-+}_0(q,\omega)$ is approximately equal to
$S^{+-}_0(q,\omega)/5$ near the low-energy threshold,
$\omega\gtrsim\varepsilon\biglb(q\mp\pi M)\bigrb)$ and
$q\approx\pm\pi M$.
That $S^{-+}_0(q,\omega)$ is much smaller than $S^{+-}_0(q,\omega)$ is 
consistent with the result in the strong-coupling limit where the
$t_{i,0}$ bosons are absent in the ground state
[see Eqs.\ (\ref{S^+_0,i}) and (\ref{S^-_0,i})].

\section{Discussion}

We shall discuss implications of the results we obtained for the
Heisenberg ladder to the $\pi$-resonance mode in the SO(5) symmetric
ladder model. 
As pointed out in Ref.~\onlinecite{Scalapino}, there is an analogy
between the quantum phase transition driven by the chemical potential
in the SO(5) symmetric ladder model and the field-induced phase
transition in the Heisenberg ladder.
Obviously, the chemical potential plays the role of the magnetic
field.
The analogy is most clearly seen in the strong-coupling
limit.\cite{Scalapino} 
At half filling the ground state of the $E_0$ phase or the Mott
insulating phase discussed in Ref.\ \onlinecite{Scalapino} is a state
in which all the rungs are in the spin singlet state.
When the chemical potential is zero, there are fivefold degenerate
low-lying massive modes above the ground state.
The five modes consist of a $S=1$ magnon triplet, a hole pair state
where two holes are placed on a single rung, and a state where two
additional electrons are put on a rung.
When the chemical potential is turned on, the energy of the hole-pair
(electron-pair) excitation decreases (increases) while the magnon
triplet is not directly affected by the chemical potential.
Thus, we see that the hole pair corresponds to the
$S^z=1$ magnon or the $t_{i,+}$ boson in the Heisenberg ladder.
The triplet magnon in the SO(5) model is an analog of the $S^z=0$
magnon ($t_{i,0}$ boson) in the Heisenberg ladder.
Furthermore, the low-energy effective Hamiltonian for the hole-pair
excitations in the strong-coupling limit is similar to the effective
Hamiltonian for the $t_{i,+}$ bosons.
That is, hole pairs may be viewed as hard-core bosons which repel each
other when two hole pairs sit on neighboring
rungs.\cite{Scalapino,Eder2} 
Let us find operators playing the role of the spin operators in the
Heisenberg ladder.
First, the operator corresponding to $S^z_{\pi,i}$ should change a
singlet rung
$|\Omega\rangle=(1/\sqrt{2})
(c^\dagger_{i,\uparrow}d^\dagger_{i,\downarrow}
 -c^\dagger_{i,\downarrow}d^\dagger_{i,\uparrow})|0\rangle$
into a hole pair or $|0\rangle$.
Here $c^\dagger_{i,\sigma}$ and $d^\dagger_{i,\sigma}$ are creation
operators of an electron with spin $\sigma$ on the $i$th rung of upper
($c$) and lower ($d$) chains.
Obviously the $d$-wave pair operator $\Delta_i$ is such an operator:
$\Delta_i=
 (c_{i,\uparrow}d_{i,\downarrow}
  -c_{i,\downarrow}d_{i,\uparrow})/\sqrt{2}$.
Second, the operator corresponding to
$S^z_{0,i}\approx t^\dagger_{i,+}t_{i,+}$ should be a number
operator of hole pairs.
It is given by
$N_i\equiv 1-
(1/2)\sum_\sigma(c^\dagger_{i,\sigma}c_{i,\sigma}
    +d^\dagger_{i,\sigma}d_{i,\sigma})$.
Finally, from the relation
$(S^+_{i,c}-S^+_{i,d})|\Omega\rangle\equiv
 (c^\dagger_{i,\uparrow}c_{i,\downarrow}
  -d^\dagger_{i,\uparrow}d_{i,\downarrow})|\Omega\rangle
 =-\sqrt{2}c^\dagger_{i,\uparrow}d^\dagger_{i,\uparrow}|0\rangle$,
we find that $\mbox{\boldmath$S$}_{i,c}-\mbox{\boldmath$S$}_{i,d}$
creates a triplet magnon from a rung singlet.
Thus we conclude that the spin operator
$\mbox{\boldmath$S$}_{i,c}-\mbox{\boldmath$S$}_{i,d}$
is an analog of $S^z_{\pi,i}$.

When the chemical potential is increased beyond the charge gap which
equals the spin gap in the SO(5) symmetric model, the ladder is doped
with the charge carrier (holes) and becomes superconducting with the
$d$-wave-like symmetry.
For spatial dimension greater than or equal to two, the superconducting
order is long-ranged in the ground state, and this gives rise to a
$\delta$-function peak or the $\pi$-resonance in the dynamic spin
structure factor.
In one dimension, however, the order is quasi-long-ranged, and
therefore the peak is expected to be replaced by a power-law
singularity.\cite{Scalapino,Lin}
The threshold energy at $q=\pi$ is also shown to be equal to the
chemical potential.\cite{Scalapino}
These features are readily reproduced from our results for the
Heisenberg ladder model.

From the approximate mapping we discussed above, the $d$-wave pair
correlation function is expected to show the quasi-long-range order
corresponding to the $XY$ order in the Heisenberg ladder,
Eq.\ (\ref{static<S^x_piS^x_pi>}):
\begin{equation}
\langle\Delta^\dagger_i\Delta_j\rangle\propto
\frac{1}{|i-j|^{1/\tilde\eta}},
\label{<DeltaDelta>}
\end{equation}
where the exponent $\tilde\eta$ is presumably smaller than 2 and
approaches 2 in the limit where the hole density $\delta$ goes to
zero.
The correlator has no $(-1)^{i-j}$ factor because the hole-pair mode
has a minimum energy at $q=0$.
The charge density correlation is related to
$\langle S^z_{0,i}S^z_{0,j}\rangle$, Eq.\ (\ref{static<S^z_0S^z_0>}),
and is also quasi-long ranged:
\begin{equation}
\langle N_iN_j\rangle-\delta^2\propto
\frac{\cos(2\pi\delta|i-j|)}{|i-j|^{\tilde\eta}}
=\frac{\cos(4k_F|i-j|)}{|i-j|^{\tilde\eta}},
\label{<NN>}
\end{equation}
where we have used the relation between the hole density and the Fermi
wave number, $\pi(1-\delta)=2k_F$.
The result that the correlations of the $d$-wave superconductivity and
$4k_F$ charge density wave show power-law decay with the exponents
whose product is 1 was also obtained by Nagaosa for a generic
two-chain model.\cite{Nagaosa95} 
Finally the spin correlation of the SO(5) ladder is expected to be
\begin{equation}
\langle(S^\alpha_{i,c}-S^\alpha_{i,d})
       (S^\alpha_{j,c}-S^\alpha_{j,d})\rangle
\propto
(-1)^{i-j}\cos(\pi\delta|i-j|)\frac{K_0(|i-j|/\xi)}{|i-j|^{\tilde\eta/4}},
\label{<SS>}
\end{equation}
where $\xi$ is the correlation length determined by the spin gap.
The spin structure factor is then
\begin{equation}
S(q,\omega)\propto
[\omega-\tilde\varepsilon(q-\pi\pm\pi\delta)]^{-1+\tilde\eta/4},
\label{structurefactor}
\end{equation}
where $\tilde\varepsilon(q)$ is the magnon dispersion at $\delta=0$.
From Fig.~\ref{fig:S^zz_pi(q,omega)} we see that the threshold energy
at $q=\pi$ is determined by the chemical potential, as expected.
Although the exponent $\tilde\eta$ depends on the detail of the model,
we can generally conclude that it is $2$ in the low-density limit of
holes ($\delta\to0$), where we may regard the hard-core bosons as
free fermions ($g=1$).
This universal exponent was independently found by Ivanov and
Lee\cite{Ivanov} and by Schulz\cite{Schulz98} for the $t$-$J$ ladder
and was also obtained by Konik {\it et al.} for
the SO(8) Gross-Neveu model.\cite{Konik} 
We thus find that the spin structure factor has a universal square-root
divergence at the critical point.
When the superfluid density is finite, the interaction between bosons
becomes important and modify the exponent, as we saw in the Heisenberg
ladder model.
We expect that the square-root singularity is a universal feature for
the spectral weight of a gapped excitation generated by injecting a
massive particle to a superfluid in the low-density limit.

Although our argument above is based on the analogy and approximate
mapping, our results should be valid as long as the weak-coupling and
strong-coupling limits are in the same phase.
We notice that Eq.\ (\ref{<SS>}) is the same as the ^^ ^^ mean-field''
result given in Sec.\ VII of Ref.\ \onlinecite{Lin}.
The validity of this result is, however, questioned by Lin {\it et
  al.} as it misses the existence of the bound states such as
the Cooper pair-magnon bound states found in the SO(8) Gross-Neveu
model.
On the other hand, we didn't find such a bound state in our
weak-coupling calculation.
It is not clear at the moment whether this is due to the approximation
we made, for example, concerning the interaction term in
${\cal H}_\perp$.
It was shown by Damle and Sachdev \cite{Damle} that this term
can indeed lead to a bound state of two magnons when $H=0$.
The fate of the bound state in the gapless phase is an open question.

Finally we conclude this paper by summarizing our results on the
spin correlations in the gapless phase of the two-leg Heisenberg ladder.
We have obtained the dynamical spin-spin correlation functions and the
structure factors, extending the bosonization theory of Shelton {\it et
al.} to the gapless regime.
The correlation functions are classified into two categories:
algebraically decaying ones, $\langle S^z_0(x,\tau)S^z_0(0,0)\rangle$
and $\langle S^\pm_\pi(x,\tau)S^\mp_\pi(0,0)\rangle$, and exponentially
decaying ones, $\langle S^z_\pi(x,\tau)S^z_\pi(0,0)\rangle$ and
$\langle S^\pm_0(x,\tau)S^\mp_0(0,0)\rangle$.
We have also found that the terms $\propto\cos(2\pi lMx)$ ($l$: integer)
are quasi-long-ranged, while the terms $\propto\cos[(2l+1)\pi Mx]$ are
short-ranged.
The exponents of the correlation functions are controlled by the single
parameter $g$, which is smaller (larger) than 1 for $J_\perp>0$
($J_\perp<0$).
The parameter $g$ approaches 1 in the limits $M\to0,1$.
The structure factors have power-law singularities at the lower edges, and
the strongest divergence is at $\omega=\pm\tilde v(q-\pi)$ in
$S^{\pm\mp}_\pi(q,\omega)$ due to the dominant $XY$ spin correlation.
The next strongest singularity is found at the lower edge of
$S^{zz}_\pi(q,\omega)$: $\omega=\{v^2[q-\pi(1\pm M)]^2+m^2\}^{1/2}$.
The exponent is universally given by $-1/2$ in the limits $M\to0,1$.

\acknowledgements

We thank I.\ Affleck, L.\ Balents, E.\ Demler, M.\ P.\ A.\ Fisher,
W.\ Hanke, H.\ H.\ Lin, M.\ Oshikawa, and D.\ Scalapino for useful
discussions. 
A.F.\ is supported by the Monbusho overseas research grant.
This work is supported by NSF under grant numbers DMR-9400372 and
DMR-9522915.

\appendix
\section*{Integrals}
In this appendix we list integral formulas we used to calculate the
dynamical structure factors.

For the gapless modes, we need the following integral:
\begin{eqnarray}
&&
\int^\infty_{-\infty}dx\int^\infty_{-\infty}dt
\frac{e^{-iqx+i\omega t}}
     {(x+\tilde vt-i0^+)^{\gamma_+}(x-\tilde vt+i0^+)^{\gamma_-}}
\nonumber\\
&&\qquad
=\Theta(\omega-\tilde vq)\Theta(\omega+\tilde vq)
\frac{2\pi^2e^{i\pi(\gamma_+-\gamma_-)/2}}
     {\tilde v\Gamma(\gamma_+)\Gamma(\gamma_-)}
\left(\frac{2\tilde v}{\omega-\tilde vq}\right)^{1-\gamma_+}
\left(\frac{2\tilde v}{\omega+\tilde vq}\right)^{1-\gamma_-}.
\label{formula1}
\end{eqnarray}

For the structure factor $S^{zz}_\pi(q,\omega)$ we first take the Fourier
transform of the correlation function in the imaginary time:
\begin{eqnarray}
I^{zz}_\pi(q,i\tilde\omega)&\equiv&
\int^\infty_{-\infty}dx\int^\infty_{-\infty}d\tau
\frac{K_0(m\sqrt{x^2+v^2\tau^2}/v)}
     {(x^2+\tilde v^2\tau^2)^\gamma}
e^{-iqx+i\tilde\omega\tau}
\nonumber\\
&=&
\frac{v}{(2\tilde v)^{2\gamma-1}}\frac{\Gamma(1-\gamma)}{\Gamma(\gamma)}
\int^\infty_{-\infty}dk\int^\infty_{-\infty}d\nu
\frac{(\nu^2+\tilde v^2k^2)^{\gamma-1}}
     {(\omega-\nu)^2+\varepsilon^2(q-k)}
\nonumber\\
&=&
\frac{v}{(2\tilde v)^{2\gamma-1}}\frac{\Gamma(1-\gamma)}{\Gamma(\gamma)}
\int^\infty_{-\infty}dk
\left(
\frac{\pi}
{\varepsilon(q-k)
 \{[\tilde\omega+i\varepsilon(q-k)]^2+\tilde v^2k^2\}^{1-\gamma}}
+\int^\infty_{\tilde v|k|}d\nu
\frac{2\sin(\pi\gamma)(\nu^2-\tilde v^2k^2)^{\gamma-1}}
     {(\tilde\omega-i\nu)^2+\varepsilon^2(q-k)}
\right).
\nonumber\\
&&
\label{convolution}
\end{eqnarray}
After the analytic continuation we take the imaginary part to find
\begin{equation}
{\rm Im}I^{zz}_\pi(q,\omega+i0^+)=
\frac{v}{(2\tilde v)^{2\gamma-1}}
\left(\frac{\pi}{\Gamma(\gamma)}\right)^2
\int^\infty_{-\infty}dk
\frac{\Theta\biglb(\omega-\varepsilon(q-k)-\tilde v|k|\bigrb)}
 {\varepsilon(q-k)
  \{[\omega-\varepsilon(q-k)]^2-\tilde v^2k^2\}^{1-\gamma}}
\label{formula2}
\end{equation}
for $\omega>0$.
When $v|q|\ll m\tilde v/v$,
$\varepsilon(q-k)+\tilde v|k|\approx \varepsilon(q)+\tilde v|k|$.
In this case we may approximate the last integral as
\begin{equation}
\frac{\Theta\biglb(\omega-\varepsilon(q)\bigrb)}{m}
\int dk
\frac{\Theta\biglb(\omega-\varepsilon(q)-\tilde v|k|\bigrb)}
     {\{[\omega-\varepsilon(q)]^2-\tilde v^2k^2\}^{1-\gamma}}
=
\Theta\biglb(\omega-\varepsilon(q)\bigrb)
\frac{B(\gamma,1/2)}{\tilde vm}
[\omega-\varepsilon(q)]^{2\gamma-1},
\label{formula3}
\end{equation}
where $B(a,b)$ is the beta function.
From Eqs.\ (\ref{formula2}) and (\ref{formula3}) we finally obtain
\begin{equation}
{\rm Im}I^{zz}_\pi(q,\omega+i0^+)=
\Theta\biglb(\omega-\varepsilon(q)\bigrb)
\frac{\pi^2}{\Gamma(2\gamma)}\frac{v}{m\tilde v}
\left(\frac{\tilde v}{\omega-\varepsilon(q)}\right)^{1-2\gamma},
\label{formula4}
\end{equation}
which is valid for $\omega-\varepsilon(q)\ll m$.

We next consider $S^{\pm\mp}_0(q,\omega)$.
According to Eq.\ (\ref{<S^+_piS^-_pi>a}), we need the following Fourier
transform: 
\begin{eqnarray}
I^{+-}_0(q,i\tilde\omega)&\equiv&
-i\int^\infty_{-\infty}dx\int^\infty_{-\infty}d\tau
\frac{K_0(m\sqrt{x^2+v^2\tau^2}/v)}
  {(x^2+\tilde v^2\tau^2)^\gamma(x-i\tilde v\tau)}
e^{-iqx+i\tilde\omega\tau}
\nonumber\\
&=&
i\frac{\Gamma(1-\gamma)}{\Gamma(1+\gamma)}
\frac{v}{(2\tilde v)^{2\gamma}}
\int^\infty_{-\infty}dk\int^\infty_{-\infty}d\nu
\frac{(\nu+i\tilde vk)^\gamma(\nu-i\tilde vk)^{\gamma-1}}
  {(\omega-\nu)^2+\varepsilon^2(q-k)}
\nonumber\\
&=&
\frac{\Gamma(1-\gamma)}{\Gamma(1+\gamma)}
\frac{v}{(2\tilde v)^{2\gamma}}\int^\infty_{-\infty}dk
\left(
\frac{i\pi[\omega+i\tilde vk+i\varepsilon(q-k)]^\gamma}
  {\varepsilon(q-k)[\omega-i\tilde vk+i\varepsilon(q-k)]^{1-\gamma}}
-2\sin(\pi\gamma)\int^\infty_{\tilde v|k|}d\nu
\frac{(\nu+\tilde vk)^\gamma(\nu-\tilde vk)^{\gamma-1}}
  {(\omega-i\nu)^2+\varepsilon^2(q-k)}
\right).
\nonumber\\
&&
\label{formula5}
\end{eqnarray}
After the analytic continuation we obtain
\begin{equation}
{\rm Im}I^{+-}_0(q,\omega+i0^+)=
\frac{v}{\gamma(2\tilde v)^{2\gamma}}
\left(\frac{\pi}{\Gamma(\gamma)}\right)^2
\int^\infty_{-\infty}dk
\Theta\biglb(\omega-\varepsilon(q-k)-\tilde v|k|\bigrb)
\frac{[\omega-\varepsilon(q-k)-\tilde vk]^\gamma}
     {\varepsilon(q-k)[\omega-\varepsilon(q-k)+\tilde vk]^{1-\gamma}}.
\label{formula6}
\end{equation}
Using the same approximation as in Eq.\ (\ref{formula3}), we obtain
\begin{equation}
{\rm Im}I^{+-}_0(q,\omega+i0^+)=
\Theta\biglb(\omega-\varepsilon(q)\bigrb)
\frac{\pi^2}{\Gamma(1+2\gamma)}\frac{v}{m\tilde v}
\left(\frac{\omega-\varepsilon(q)}{\tilde v}\right)^{2\gamma}
\label{formula7}
\end{equation}
for $0<\omega-\varepsilon(q)\ll m$ and $v|q|\ll m\tilde v/v$.
In the same way we get
\begin{eqnarray}
{\rm Im}I^{-+}_0(q,\omega+i0^+)&\equiv&
{\rm Im}\lim_{i\tilde\omega\to\omega+i0^+}
i\int^\infty_{-\infty}dx\int^\infty_{-\infty}d\tau
\frac{K_0(m\sqrt{x^2+v^2\tau^2}/v)}
     {(x^2+\tilde v^2\tau^2)^\gamma(x+i\tilde v\tau)}
e^{-iqx+i\tilde\omega\tau}
\nonumber\\
&=&
\frac{v}{\gamma(2\tilde v)^{2\gamma}}
\left(\frac{\pi}{\Gamma(\gamma)}\right)^2
\int^\infty_{-\infty}dk
\Theta\biglb(\omega-\varepsilon(q-k)-\tilde v|k|\bigrb)
\frac{[\omega-\varepsilon(q-k)+\tilde vk]^\gamma}
   {\varepsilon(q-k)[\omega-\varepsilon(q-k)-\tilde vk]^{1-\gamma}},
\label{formula8}
\end{eqnarray}
which reduces to Eq.\ (\ref{formula7}) for $0<\omega-\varepsilon(q)\ll m$
and $v|q|\ll m\tilde v/v$.

\begin{figure}
\begin{center}
\leavevmode\epsfxsize=65mm
\epsfbox{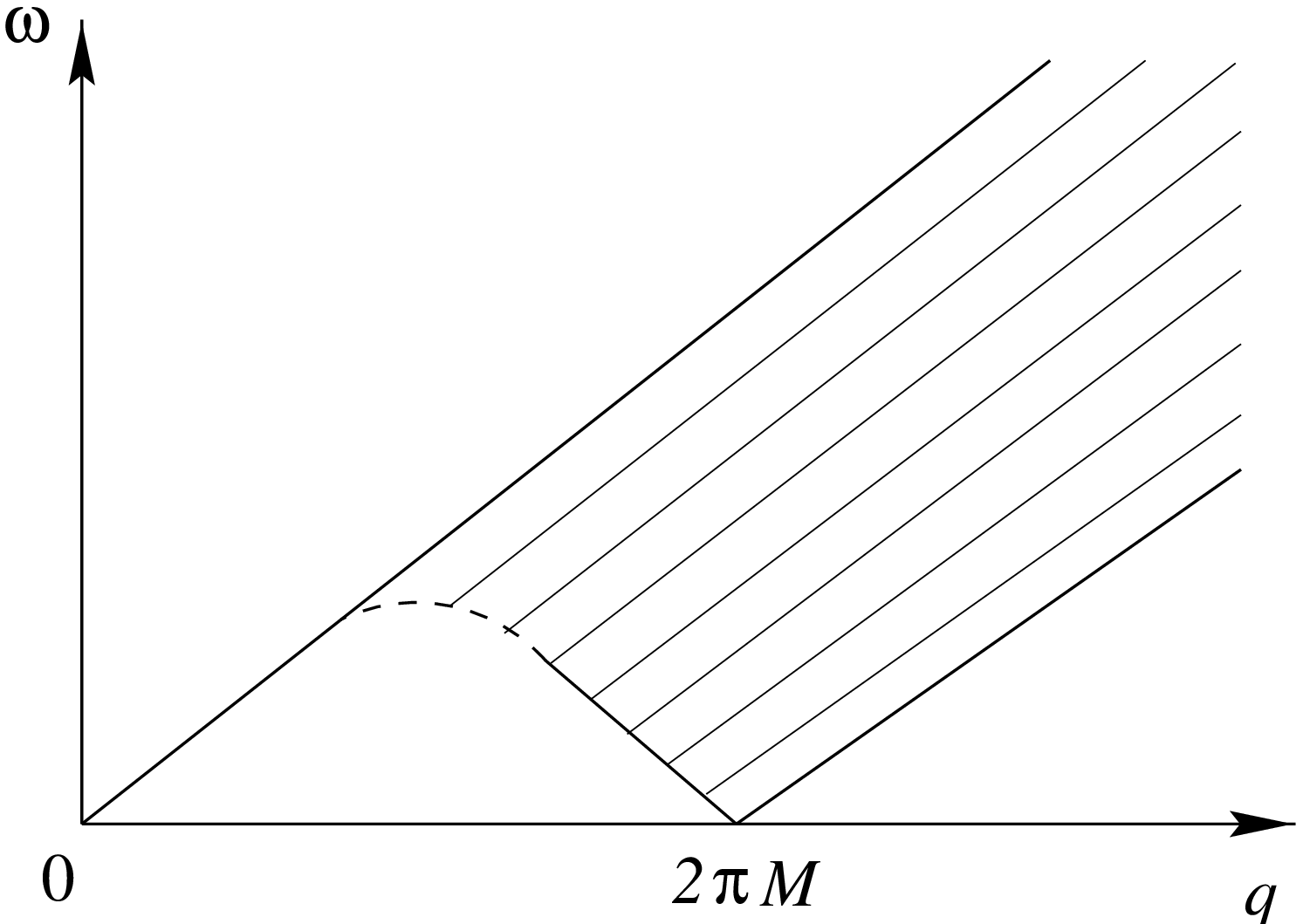}
\end{center}
\caption{Support of $S^{zz}_0(q,\omega)$.
The shaded region shows where $S^{zz}_0(q,\omega)$ is nonzero.}
\label{fig:S^zz_0(q,omega)}
\end{figure}

\begin{figure}
\begin{center}
\leavevmode\epsfxsize=65mm
\epsfbox{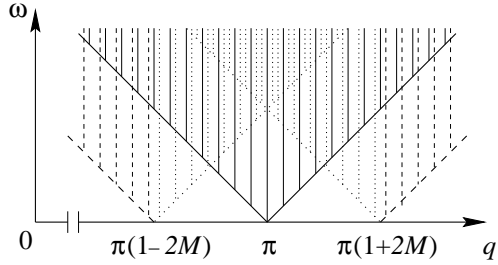}
\end{center}
\caption{Support of $S^{+-}_\pi(q,\omega)$.
The shaded regions show where $S^{+-}_\pi(q,\omega)$ is nonzero.
The strongest divergence is at $\omega=\pm\tilde v(q-\pi)$.
The next strongest singularity is at $\omega=\pm\tilde v[q-\pi(1\pm2M)]$.
}
\label{fig:S^+-_pi(q,omega)}
\end{figure}

\begin{figure}
\begin{center}
\leavevmode\epsfxsize=65mm
\epsfbox{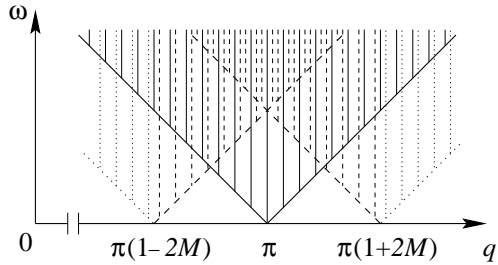}
\end{center}
\caption{Support of $S^{-+}_\pi(q,\omega)$.
The shaded regions show where $S^{-+}_\pi(q,\omega)$ is nonzero.
The strongest divergence is at $\omega=\pm\tilde v(q-\pi)$.
The next strongest singularity is at $\omega=\mp\tilde v[q-\pi(1\pm2M)]$.
Note the difference from Fig.~{\protect\ref{fig:S^+-_pi(q,omega)}}.}
\label{fig:S^-+_pi(q,omega)}
\end{figure}

\begin{figure}
\begin{center}
\leavevmode\epsfxsize=65mm
\epsfbox{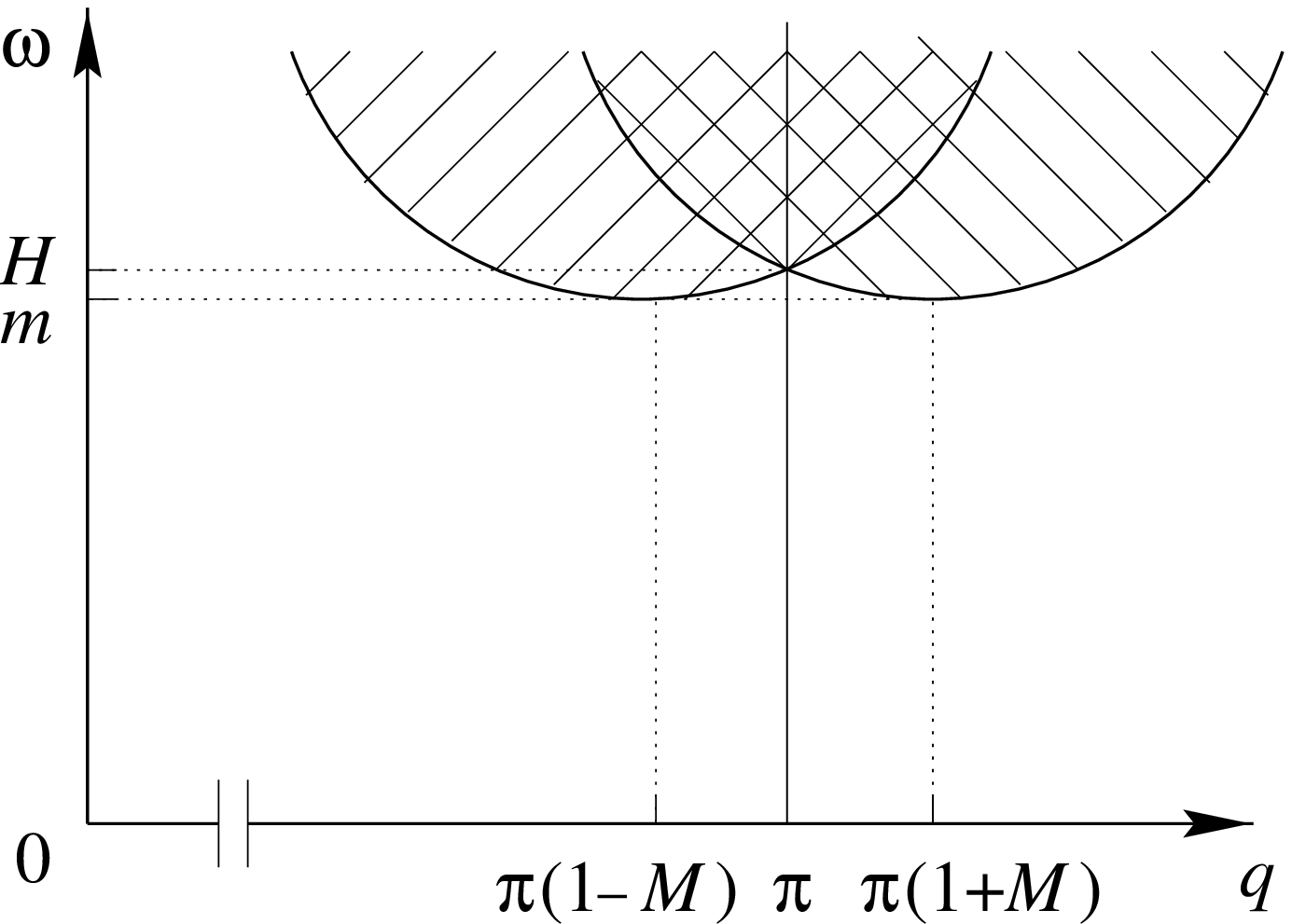}
\end{center}
\caption{Support of $S^{zz}_\pi(q,\omega)$.
The shaded regions show where $S^{zz}_\pi(q,\omega)$ is nonzero.}
\label{fig:S^zz_pi(q,omega)}
\end{figure}

\begin{figure}
\begin{center}
\leavevmode\epsfxsize=65mm
\epsfbox{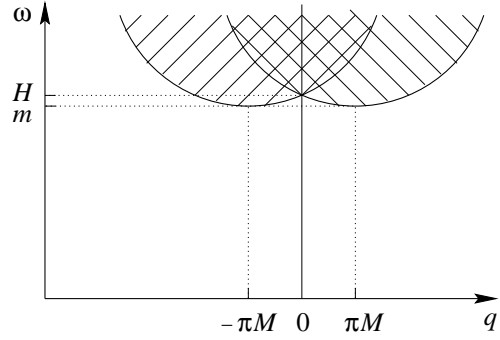}
\end{center}
\caption{Support of $S^{+-}_0(q,\omega)$ and $S^{-+}_0(q,\omega)$.
The shaded regions show where $S^{\pm\mp}_0(q,\omega)$ are nonzero.}
\label{fig:S^+-_0(q,omega)}
\end{figure}

\end{document}